\documentstyle[psfig]{l-aa}

\def\cgs{{erg cm$^{-2}$ s$^{-1}$}}

\begin{document}
\thesaurus{11.17.3; 11.17.2; 13.25.2)}   
\title{BeppoSAX broad-band observations of low-redshift quasars: Spectral                                                    
curvature and iron K$_{\alpha}$ lines}

\author{T. Mineo\inst{1}, 
F. Fiore\inst{2,3,4}, 
A. Laor\inst{5},
E. Costantini\inst{4,6},
W.N. Brandt\inst{7},
A. Comastri\inst{6},
R. Della Ceca\inst{8},
M. Elvis\inst{4}, 
T. Maccacaro\inst{8},
S. Molendi\inst{9}
}
\institute{Istituto di Fisica Cosmica ed Applicazioni
all'Informatica CNR, Via U. La Malfa 153, I-90146, Palermo, Italy \and 
BeppoSAX Science Data Center, c/o Telespazio, Via Corcolle 19,
I-00131 Roma, Italy \and
Osservatorio Astronomico di Roma, I-00040, Monteporzio (Rm), Italy \and
Harvard Smithsonian Center for Astrophysics, Cambridge, MA, USA \and
Physics Department, Technion, Haifa 32000, Israel \and
Osservatorio Astronomico di Bologna, Via Ranzani 1, I-40127, Bologna, Italy \and
Department of Astronomy and Astrophysics, The Pennsylvania State University,
525 Davey Lab, University Park, PA 16802 USA \and
Osservatorio Astronomico di Brera, Via Brera 28, I-20121, Milano, Italy \and
Istituto di Fisica Cosmica e Tecnologia Relative CNR, Via Bassini 15, I-20133, Milano Italy 
}

\offprints{T. Mineo: mineo@ifcai.pa.cnr.it}
\date{Received ....; accepted ....}
\maketitle

\markboth{T. Mineo et al.: BeppoSAX broad band observations of low redshift
quasars}
{T. Mineo et al.: BeppoSAX broad band observations of low redshift
quasars}

\begin{abstract}

We present results from BeppoSAX observations of 10 low redshift
quasars. The quasars are part of the Laor et al. (1997) sample of 23
optically selected PG quasars with redshift $<0.4$ and low Galactic
absorption along the line of sight. Significant spectral curvature is 
detected for the 6 quasars with the highest signal to noise ratio
in their low energy spectra. The average curvature can be parameterized 
as a flattening of the underlying power law by $\Delta\alpha\sim0.5$
above $\sim$ 1 keV.  
We find that quasars with a steeper soft X-ray (0.1--2
keV) spectrum tend to be steeper also at higher (2--10 keV) energies.
The distribution of the best fit 2--10 keV slopes is similar
to that found for other radio-quiet AGN  
and characterized by a large dispersion around the mean: $\alpha_E 
\simeq$ 1.0 $\pm$ 0.3.
Iron K$_{\alpha}$ lines are detected in 4 quasars. 
In the narrow--line quasar PG~1115$+$407, the rest frame line energy 
(6.69$\pm$0.11 keV) and equivalent width (580$\pm$280 eV) are consistent
with those found in
a few low redshift narrow--line Seyfert 1 galaxies (NLSy1). This, together
with the similarity of the 0.1--10~keV X-ray  continuum, suggests that 
this quasar is the higher redshift and luminosity analog of a NLSy1. 
In the broad line quasar PG~0947$+$396, the rest-frame line energy
suggests fluorescence from cold iron. The line equivalent width 
($>$ 400 eV) is however about 2--3 times higher than that usually found in 
Seyfert 1 galaxies. 
The high energy power-law slopes and the iron line 
emission properties seem to be unrelated to the X--ray luminosity.

\end{abstract}

\maketitle

\section{Introduction}

Observations of quasars above 2 keV have in the past been largely
devoted to objects selected via  X-rays. Their 2--10 keV spectral
indices cluster tightly around a ``canonical'' value of 0.9 
($f_E \propto E^{-\alpha}$; Williams 
et al. 1992, Comastri et al. 1992, Reeves et al. 1997, Lawson \&
Turner 1997).  This uniformity may be the result of a selection effect
(Elvis 1992). Seyfert galaxies, in fact, do indeed show a significantly
wider range of 2--10 keV spectral indices (Brandt, Mathur \& Elvis 1997) and
QSOs might show the same effect as reported by recent studies 
of ASCA observations (George et al. 2000).
 We have therefore started a program to observe, with BeppoSAX 
(Boella et al. 1997a), a reasonably
large, well defined, and representative sample of optically selected
quasars. Our goals are to perform a systematic study of their hard ($E>2$ keV)
emission spectra and, taking advantage of the broad BeppoSAX band, to
compare them with the soft 0.1--2 keV spectra acquired simultaneously. 
 
BeppoSAX has so far observed 10 quasars  extracted from the
Laor et al. (1997) sample, nine as part of a Core Program and one as part
of the Science Verification Phase (SVP).  The Laor et al. sample consists of
23 PG quasars selected to have $z<0.4$, Galactic
$N_{\rm H}<1.9 \times10^{20}$ cm$^{-2}$ and $M_B<-23$.  These quasars all have
good ROSAT PSPC X-ray spectra, radio fluxes, IR photometry, high
S/N optical spectro photometry, IUE and HST spectra (Laor et al.
1997 and references therein).  

The 0.2--2 keV PSPC spectra of most of
these 23 quasars are well fitted by a single power law plus Galactic
absorption. The average spectral index is $\alpha_E=1.62\pm0.45$, and the
$\alpha_E$ range is 0.9--2.8.  The large spread in $\alpha_E$
permitted the discovery of  correlations between the soft
X-ray spectral shape and optical emission line properties (Laor et al. 1994,
1997, Boller, Brandt \& Fink 1996, Ulrich-Demoulin \& Molendi 1996): 
steeper soft X-ray quasars tend to have narrower $H_\beta$
lines, fainter [OIII] emission and stronger FeII emission with respect to
$H_\beta$.  Laor et al. (1997) interpret the $\alpha_E-H_\beta$ FWHM
anticorrelation in terms of a dependence of $\alpha_E$ on $L/L_{\rm
Edd}$. The line width is inversely proportional to $\sqrt {L/L_{\rm
Edd}}$ if the broad line region is virialized and if its size is
determined by the central source luminosity (see Laor et al. 1997, \S
4.7).  So narrow-line, steep (0.1--2 keV) spectrum AGNs emit close to
the Eddington luminosity (Nicastro  2000).  

A proposed scenario for these sources,
as described by Pounds et al. (1995), is that the hard X-ray power law
is produced by Comptonization in a hot corona: as the object becomes
more luminous in the optical-UV, Compton cooling of the corona
increases. The corona becomes colder, thus producing a steeper X-ray
power law.  If this mechanism is operating, then steep
$\alpha_E$(PSPC) quasars should also have a steep hard X-ray
power law  that would have been missed by  hard X-ray
surveys.  BeppoSAX 0.1--10 keV observations should help in verifying
this picture; test if the correlations between the soft X-ray spectrum
and the optical lines properties hold also at high X-ray energies; and 
study the Iron K emission lines in high luminosity objects, to test the
claim of an X-ray Baldwin effect (Iwasawa \& Taniguchi 1993, Nandra et al. 1997).

In this paper we report the analysis of BeppoSAX 
observations of 10 PG quasars.
In \S 2 we describe the observations and the data
analysis; in \S 3 we present the broad band 0.1--100 keV spectra;
in \S 4 we discuss in detail the 0.1--10 keV spectral shape and compare
it with the PSPC results; in \S 5 we present results on the iron line;
in \S 6 we  present our conclusions.
A correlation between the 0.1--10 keV spectral shape and the
optical line properties is deferred to a future 
publication.

In the paper
$H_{0}$=50 km s$^{-1}$ Mpc$^{-1}$ and $q_{0}$=0.5 are assumed.

\section{Observations and Data Reduction}

Table~\ref{tab1} lists the PG quasars observed by BeppoSAX, together with
their redshifts (Schmidt \& Green 1983) and 21~cm Galactic column 
densities (we use the values quoted in
Laor et al. 1997, where the relevant references are also reported).
PG~1226$+$023 (3C 273)  was observed during 
the SVP, and  a detailed wide band spectral analysis 
has been published by Grandi et al. (1997).

\begin{table}
\setlength{\tabcolsep}{0.4pc}
\catcode`?=\active \def?{\kern\digitwidth}
\caption{PG~quasars observed by BeppoSAX}
\label{tab1}
\begin{center}
\begin{tabular}{llcc}
\hline
\multicolumn{1}{c} {Quasar} &
\multicolumn{1}{c} {Other Name} &
\multicolumn{1}{c} {$z$} &
\multicolumn{1}{c} {$N_{\rm H}$}  \\ 
  &  &   & $(10^{20}$ cm$^{-2})$ \\
\hline
PG~0947$+$396 & K 347-45 & 0.206 & 1.92  \\
PG~1048$+$342 &          & 0.167 & 1.74  \\
PG~1115$+$407 &          & 0.154 & 1.74  \\
PG~1202$+$281 & GQ Comae & 0.165 & 1.72  \\
PG~1226$+$023 & 3C 273   & 0.158 & 1.68  \\
PG~1352$+$183 & PB 4142  & 0.158 & 1.84 \\
PG~1402$+$261 & TON 182  & 0.164 & 1.42 \\
PG~1415$+$451 &          & 0.114 & 0.96  \\
PG~1512$+$370 & 4C 37.43 & 0.371 & 1.40 \\
PG~1626$+$554 &          & 0.133 & 1.82 \\
\hline
\end{tabular}
\end{center}
\end{table}

All
quasars but two are radio quiet according to the  
definition given by Kellermann et al. (1989)  based on the ratio $R$ between  
the radio flux at 6 cm and the  optical
flux  at 4400 \AA: radio-loud quasars present  $R>10$.
This criterion corresponds to selecting $\alpha_{ro} \leq 0.2$ where $\alpha_{ro}$
is the radio--to--optical spectral index defined as 
$log(S(5 GHz)/S(2500 \AA))/5.38$ (Stocke et al. 1985).
The radio-loud quasars in
our sample are PG~1512$+$370 ($R=190$; Kellermann et al. 1989) and PG~1226$+$023
($R=1138$; Kellermann et al. 1989). 
\\
Table~\ref{tab2} shows the  dates, the observing codes and the exposure times
for the BeppoSAX observations of the ten PG quasars.
These observations were performed with the  Narrow Field
Instruments, LECS (0.1--10 keV; Parmar et al. 1997), MECS (1.3--10
keV, Boella et al. 1997b), HPGSPC (4--60 keV, Manzo et al. 1997) and
PDS (13--200 keV, Frontera et al. 1997).  We report here the analysis
of the LECS, MECS and PDS data; the HPGSPC has a large and structured
background which makes it not sensitive enough for faint extragalactic
sources.

At launch the MECS was composed of three identical
units. Unfortunately on 1997 May 6$^{\it th}$ a technical failure
caused the unit MECS1 to switch off. All observations after this date
were performed with two units (MECS2 and MECS3).  The MECS energy
resolution is about  8 \% at 6~keV.  The LECS is operated
during spacecraft dark time only; therefore LECS exposure times are
usually smaller than MECS ones by a factor 1.5--3. 
The PDS is a collimated instrument with a FWHM of about
1.4 degrees. Since it has a larger field of view than 
the LECS and MECS ($\sim 30$ arcmin radius) and no angular resolution,
any PDS detection has to be validated by a simultaneous fit with
LECS and MECS spectra. The PDS collimators are rocked on and off
source every 96 seconds, to simultaneously monitor the background, the
PDS on-source time is then typically half of the total integration
time.

\begin{figure*}
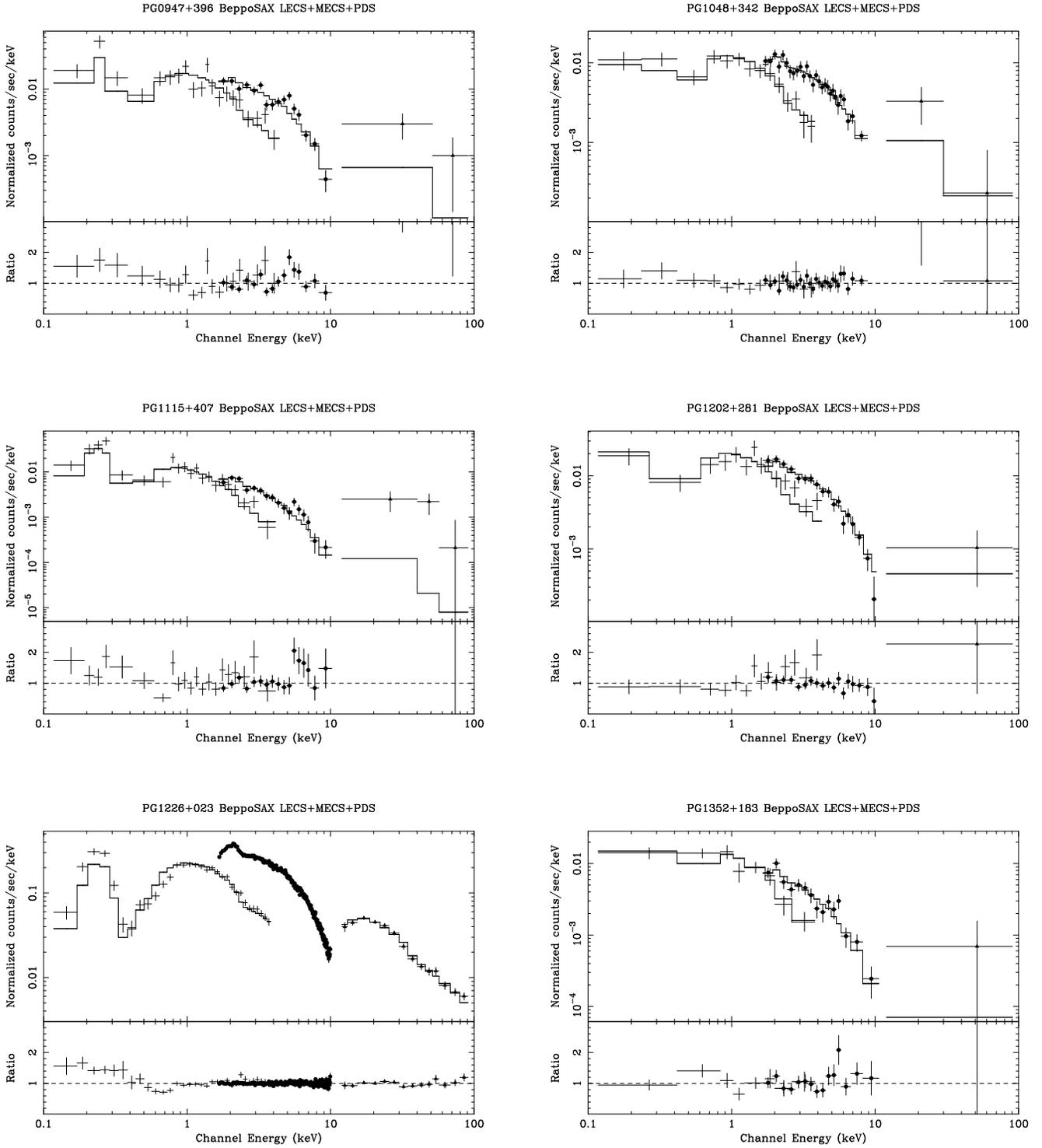

\centerline{ \vbox{
\hbox{
\psfig{figure=ms9643_a.f1,width=9.2cm,angle=-90,clip=}
\psfig{figure=ms9643_b.f1,width=9.2cm,angle=-90,clip=}
}
\hbox{
\psfig{figure=ms9643_c.f1,width=9.2cm,angle=-90,clip=}
\psfig{figure=ms9643_d.f1,width=9.2cm,angle=-90,clip=}
}
\hbox{
\psfig{figure=ms9643_e.f1,width=9.2cm,angle=-90,clip=}
\psfig{figure=ms9643_f.f1,width=9.2cm,angle=-90,clip=}
}
}}
\caption{  Joint LECS (crosses), MECS (filled circles) and PDS (triangles)
 spectra of the ten PG~quasars fitted with a single 
power law plus Galactic absorption (continuous curve). 
In each panel, the lower portion reports the data-to-model ratios.} 
\label{fig1}
\end{figure*}

\setcounter{figure}{0}
\begin{figure*}
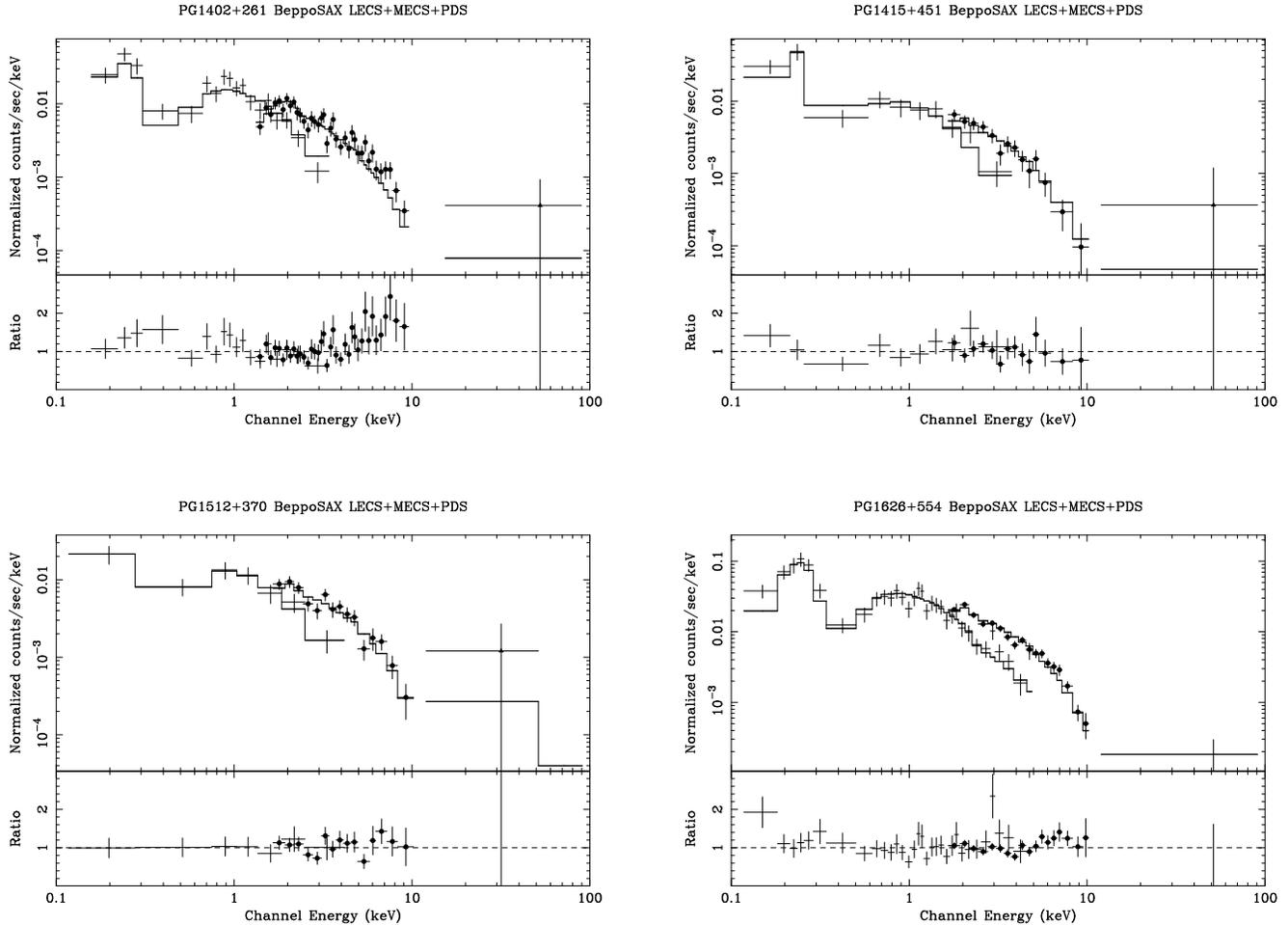

\vbox{
\hbox{ 
\psfig{figure=ms9643_g.f1,width=9.2cm,angle=-90,clip=}
\psfig{figure=ms9643_h.f1,width=9.2cm,angle=-90,clip=}
}
\hbox{
\psfig{figure=ms9643_i.f1,width=9.2cm,angle=-90,clip=}
\psfig{figure=ms9643_l.f1,width=9.2cm,angle=-90,clip=}
}
}
\caption{continued}
\label{fig1bis}
\end{figure*}

Standard data reduction was performed using the  software
package "SAXDAS".\footnote{see http://www.sdc.asi.it/software}  
In particular, data are linearized and cleaned from
Earth occultation periods and unwanted periods of high particle
background (satellite passages through the South Atlantic Anomaly).
The LECS, MECS and PDS background is relatively small and stable
(variations of at most 30 \% around the orbit) due to the 
low inclination orbit (3.95 degrees). Therefore, the data quality does not 
depend strongly on screening criteria such as Earth elevation angle,
bright Earth angle or magnetic cut-off rigidity (we accumulated data
for Earth elevation angles $>5$ degrees and magnetic cut-off rigidity
$>6$). For the PDS data we adopted an energy and temperature
dependent rise-time selection, which decreases the PDS background by
$\sim 40 \%$. This improves the signal to noise ratio of faint sources
by about a factor of 1.5 (Fiore, Guainazzi \& Grandi 1999b).

Data from the four PDS units and the 
three (or two) MECS units have been merged after equalization to produce
single MECS and PDS spectra.

We extracted spectra from the LECS and MECS using 6 arcmin and 3
arcmin radius regions respectively for all the quasars but PG
1115$+$407, PG~1226$+$023 and PG~1626$+$554.  These radii maximize the
signal-to-noise ratio below 1 keV in the LECS and above 2 keV in the
MECS.  For PG~1115$+$407 we adopted slightly smaller extraction radii
(5.3 arcmin for the LECS and 2.7 arcmin for the MECS) to reduce
the contamination  of several faint sources near the quasar
(Fiore et al.  1998, 1999a).  The cumulative intensity
of these sources is not negligible in comparison to the flux of the
quasar.  For PG~1626$+$554 we used a 5.3 arcmin radius region to extract 
the LECS spectrum in order to avoid  contamination from a source 
at $\sim$ 10 arcmin off-axis. 
The signal-to-noise ratio for PG~1226$+$023, the strongest
source in our sample, is maximum at 10.7 and 5.3 arcmin respectively
for the LECS and the MECS: these radii have been then used to
accumulate the spectra.

LECS and MECS internal backgrounds depend on the position within the field of view
(see Chiappetti et al. 1998, BeppoSAX Cookbook,\footnote{see
http://www.sdc.asi.it/software/cookbook} Parmar et al. 1998,
Fiore, Guainazzi \& Grandi 1999b).
Accordingly, background spectra were extracted from high Galactic
latitude `blank' fields from the same region of the 
detectors. We have compared the mean level of the
background in these observations with that in the quasar observations
using source free regions at various positions in the detectors:
\begin{table*}
\caption{Journal of the BeppoSAX observations}
\label{tab2}
\begin{center}
\begin{tabular}{ll@{\hspace{0.1cm}}lcr@{\hspace{0.1cm}}r@{\hspace{0.2cm}}r
l@{\hspace{0.1cm}}l@{\hspace{0.1cm}}c}
\hline
\multicolumn{1}{c} {Quasar} &
\multicolumn{2}{c} {Starting Dates} &
\multicolumn{1}{c} {Observing} &
\multicolumn{3}{c} {Exposure (s)} &
\multicolumn{3}{c} {Rate (10$^{-2}$ counts s$^{-1}$)} \\
 &
\multicolumn{2}{c} {(yy mm dd, hh:mm:ss)} &
\multicolumn{1}{c} {Code} &
\multicolumn{1}{c} {LECS}  &
\multicolumn{1}{c} {MECS}&
\multicolumn{1}{c} {PDS}&
\multicolumn{1}{c} {LECS$^{a}$}   &
\multicolumn{1}{c} {MECS$^{a}$} &
\multicolumn{1}{c} {PDS$^{b}$} \\
\hline
PG~0947$+$396 & 1997~Apr~19, &  09:39:44 & 50124001 & 10\,617               & 18\,507 
          & 8\,004 
            & 3.79$\pm$0.20              & 4.21$\pm$0.16       & 15.7$\pm$5.9\\
PG~1048$+$342 & 1997~May~3, &  09:46:38   & 50124002 & 16\,030               & 32\,633 
          & 13\,323          
            & 2.67$\pm$0.15              & 3.38$\pm$0.11       & $<$9.0 \\
PG~1115$+$407 & 1997~May~2, &  09:45:15   & 50124003 & 17\,337               & 33\,797 
          & 14\,099
            & 2.62$\pm$0.15$^{c}$        & 1.75$\pm$0.08$^{c}$ & 11.5$\pm$4.4 \\
PG~1202$+$281 & 1997~Dec~11, &  05:38:02 & 50124006 & 6\,344                & 18\,201 
          & 8\,085
            & 4.45$\pm$0.28              & 4.27$\pm$0.16       & $<$11.6 \\
PG~1226$+$023 & 1996~Jul~18, &  02:03:25 & 50021001 & 11\,637               & 129\,886
          & 60\,639
            & 61.8$\pm$0.7$^{d}$         & 131.7$\pm$0.3$^{d}$ & 158.9$\pm$3.6\\
PG~1352$+$183 & 1998~Jan~31, &  15:00:35 & 50124007 & 8\,713                & 19\,378 
          & 9\,508
            & 2.61$\pm$0.19              & 1.95$\pm$0.11       & $<$13.9\\
PG~1402$+$261 & 1999~Jan~17, & 00:18:11 & 50551002 & 13\,225              & 33\,033 
          & 14\,983
            & 3.02$\pm$0.19.             & 2.64$\pm$0.09       & $<$10.5 \\
PG~1415$+$451 & 1998~Jan~22, &  01:28:37 & 50124004 & 8\,571                & 24\,133 
          & 10\,916
            & 2.13$\pm$0.18              & 1.24$\pm$0.08       & $<$13.0 \\
PG~1512$+$370 & 1998~Jan~25, &  19:17:01 & 50124008 & 5\,101                & 18\,611 
          & 8\,805
            & 2.90$\pm$0.27              & 2.25$\pm$0.12       & $<$14.3 \\
PG~1626$+$554 & 1998~Feb~25, &  17:33:13 & 50551001 & 9\,857                & 30\,366 
          & 14\,460
            & 6.99$\pm$0.28$^{e}$        & 5.06$\pm$0.13       & $<$11.1 \\
\hline
\end{tabular}
\end{center}

$^{a}$ whenever not explicitly declared the accumulation radius is 6 arcmin for 
the LECS and 3 arcmin for the MECS \\
$^{b}$ PDS upper limits are given at the 2 $\sigma$  level \\
$^{c}$ the accumulation radius is 5.3 arcmin  for the LECS and 2.7 arcmin for the MECS \\
$^{d}$ the accumulation radius is 10.7 arcmin  for the LECS and 5.3 arcmin for the MECS \\
$^{e}$ the accumulation radius is 5.3 arcmin  \\
\end{table*}


\begin{enumerate}
\item[I)]
In the LECS the mean local background above 1 keV is consistent with the `blank'
field mean background in all cases but PG~1048$+$342 and PG~1626$+$554,
where the local background is higher by 14 and 7 \% respectively.  
The local low energy (0.1--1 keV) background is higher than the
`blank' field 0.1--1 keV background by 30--40 \% in four 
cases (PG~1048$+$342, PG~1415$+$451, PG~1202$+$281 and PG~1626$+$554). The
total excess soft background, scaled to the source extraction region,
is of the order of $10^{-3}$ counts s$^{-1}$.  
\\
In all cases but PG~1048$+$342 and PG~1626$+$554, we have not
modified the 'blank' field background spectra, but we have  compared
the significance of any feature below 1 keV with
the size of the background excess. In PG~1048$+$342 and PG~1626$+$554 
the `blank' field background spectra were scaled by the above  amount.

\item[II)]
In the MECS the local background is higher than the 'blank' field
background in seven out of ten cases (by 9 to 20 \%). In all cases
the excess was constant with energy.  We have then scaled by the same
amount the 'blank' field background spectra before subtraction.
\end{enumerate}

\noindent
Table~\ref{tab2} gives 
the detected source count rates in the energy ranges used for the spectral 
analysis for each instrument; PDS upper limits are given at the 2 $\sigma$
level. All quasars are detected in the LECS and MECS instruments.  A positive
detection (better than 2 $\sigma$ level) is  present in the PDS
for PG~0947$+$396, PG~1115$+$407 and PG~1226$+$023.

No flux variability has been detected in any of the observed
sources, except for PG~1402$+$261,  for which  a 10 \% variation in the 
MECS total count rate  is seen in time scale of the order of 15 ks.  
The variability reaches  30~\%  in the higher energy range (E$>$4 keV). 

Errors quoted in this paper represent 1 $\sigma$ uncertainty for 1
interesting parameter ($\Delta \chi^2$=1.0).

\section{0.1--100 keV spectral analysis}

Spectral fits were performed using the XSPEC 9.0 software package and
public response matrices issued in November 1998.  PG~1226$+$023
was observed at an off-axis angle of 2 arcmin, in a position close to one of
the wires supporting the LECS window. The shadowing of this wire
affects the resulting low energy spectrum and an appropriate effective
area file for this position was created using the LECS matrix
generation ``LEMAT'' code.\footnote{see
http://www.sdc.asi.it/software/saxdas/lemat.html} 
 
PI channels are rebinned sampling
the instrument resolution with the same number of channels at all
energies when possible, but with the requirement of having at least 20 counts
in each bin. This guarantees the applicability of the $\chi^2$ method in
determining the best-fit parameters. 
 
Constant factors have been introduced in the fitting models in order
to take into account the intercalibration systematic uncertainties
between instruments (BeppoSAX Cookbook, 
Fiore, Guainazzi \& Grandi 1999b).  In the fits we use the MECS
as reference instruments and constrain the LECS and PDS parameters to vary in
the range 0.7--1 and  0.77--0.95 respectively.

The energy ranges used for the fits are: 0.1--4 keV for the LECS
(channels 11--400), 1.65--10 keV for the MECS (channels 37--220) and
15--100 keV for the PDS.

LECS, MECS and PDS spectra of the PG quasars were first fitted with a single
power law of the form $f_{E}=e^{-N_{\rm H} \sigma(E)}f_{o} \, E^{-\alpha}$,
where $f_{E}$ is the flux density, $\sigma(E)$ is the absorption cross 
section per H atom (Morrison \& McCammon 1983), $f_{o}$ is the flux density at 1 keV,
$\alpha$ is the energy index and $E$ is in units of keV.   The absorbing
column
$N_{\rm H}$ has been fixed to its
Galactic value.  Figure~\ref{fig1}  shows the spectra  along
with the best fit power-law model and the residuals expressed as
data-to-model ratio.  
 Inspection of this figure allows one to note that
features are present 
in the 1--10~keV residuals of most of the spectra, as it will be discussed 
in detail
in the next section. For all quasars but PG~0947$+$396 and 
PG~1115$+$407, the PDS points are 
fully consistent with LECS--MECS extrapolations.

\begin{table*}
\caption{Results of the fit  of LECS and MECS spectra  with a single power law}
\label{tab3}
\begin{center}
\begin{tabular}{lcccccclc}
\hline
\multicolumn{1}{c} {Quasar} &
\multicolumn{1}{c} {Fit$^a$} &
\multicolumn{1}{c} {$N_{\rm H}^b$}   &
\multicolumn{1}{c} {Energy Index} &
\multicolumn{1}{c} {$F_{0.1-2}^c$} &
\multicolumn{1}{c} {$F_{2-10}^c$} &
\multicolumn{1}{c} {$L_{2-10}^d$} &
\multicolumn{1}{c} {$\chi^2_{\rm pw}$(dof)} &
\multicolumn{1}{c} { $F_{15-100}^f$}\\
\\
\hline
PG~0947$+$396 & 1 & 0.57$\pm$0.35 & 0.85$\pm$0.07 & 2.68$\pm$0.16 &
2.49$\pm$0.10 & 4.9 & 53.4 (32) &   1.58$\pm$0.59 \\
            & 2 & 1.92          & 0.97$\pm$0.07 &     &      
&        & 71.7 (33) \\
            
PG~1048$+$342 & 1 & 1.08$\pm$0.42 & 0.82$\pm$0.07 & 1.96$\pm$0.13 &
2.09$\pm$0.07 & 2.7 &  23.4 (38) &   $<$0.94  \\
            & 2 & 1.74          & 0.87$\pm$0.06 &       &       
&        & 25.6 (39) \\
            
PG~1115$+$407 & 1 & 0.34$\pm$0.32 & 1.29$\pm$0.09 & 2.72$\pm$0.16 &
0.93$\pm$0.04 & 1.0 &  40.8 (33) &   0.96$\pm$0.36 \\
            & 2 & 1.74          & 1.51$\pm$0.07 &         &       
&        & 54.2 (34) \\
            
PG~1202$+$281 & 1 & 2.40$\pm$0.69 & 0.97$\pm$0.08 & 2.77$\pm$0.22 &
2.62$\pm$0.10  & 3.3 & 27.2 (28) &   $<$1.18\\
            & 2 & 1.72          & 0.92$\pm$0.06 &         &       
&        & 28.4 (29) \\
            
PG~1226$+$023$^{e}$ & 1 & 0.88$\pm$0.10 & 0.59$\pm$0.06 & 40.55$\pm$0.68 &
69.80$\pm$0.19  & 80.0 &  301.3 (207) &   16.03$\pm$0.04\\
            & 2 & 1.68          & 0.60$\pm$0.05 &        &      
&            & 338.9 (208) \\
                       
PG~1352$+$183 & 1 & 1.96$\pm$0.61 & 1.27$\pm$0.12 & 1.98$\pm$0.15 &
1.05$\pm$0.07  & 1.2 & 22.4 (20) &   $<$1.27\\
            & 2 & 1.84          & 1.26$\pm$0.06 &       &       
&            & 22.6 (21) \\   
                 
PG~1402$+$261 & 1 & $<$0.62       & 1.24$\pm$0.08 & 1.16$\pm$0.24 &
1.53$\pm$0.10  & 1.9 & 66.9 (53) &   $<$0.89\\
            & 2 & 1.42          & 1.40$\pm$0.07 &       &       
&            & 70.4 (54) \\
            
PG~1415$+$451 & 1 & 0.52$\pm$0.49 & 1.34$\pm$0.14 & 2.09$\pm$0.25 &
0.69$\pm$0.05  & 0.4 & 16.2 (21) &   $<$1.12\\
            & 2 & 0.96          & 1.42$\pm$0.09 &       &       
&            & 16.8 (22) \\
            
PG~1512$+$370 & 1 & 1.42$\pm$0.69 & 1.12$\pm$0.12 & 1.97$\pm$0.22 &
1.23$\pm$0.07  & 8.5 & 16.7 (20) &   $<$1.37 \\
            & 2 & 1.40          & 1.12$\pm$0.10 &       &       
&        & 16.7 (21) \\
            
PG~1626$+$554 & 1 & 1.03$\pm$0.29 & 1.17$\pm$0.06 & 5.30$\pm$0.26 &
2.92$\pm$0.09  & 2.4 & 47.3 (42) &   $<$1.01\\
            & 2 & 1.82          & 1.25$\pm$0.05 &        &       
&        & 52.8 (43) \\
\hline
\end{tabular}
\end{center}

$^a$ Fit 1: $N_{\rm H}$ as free parameter; 
Fit 2: $N_{\rm H}$ fixed to the Galactic value \\
$^b ~10^{20}$ cm$^{-2}$; \hspace{1cm}
$^c$ 10$^{-12}$ \cgs; \hspace{1cm}
$^d$ 10$^{44}$ erg s$^{-1}$ \\
$^{e}$   Introducing an edge in the fitting model (2) we get 
$\chi^2$= 281.9 (206 dof); \\
 $^{f}$ 10$^{-11}$ \cgs; PDS fluxes or 2$\sigma$ upper limits obtained 
extrapolating the
LECS+MECS spectral shapes. 
\end{table*}

The PDS detections in PG~0947$+$396 and PG~1115$+$407 are well above the
LECS--MECS relative extrapolations. This excess could be due to
serendipitous sources present in the large (1\fdg4  FWHM) PDS
field of view: confusion ultimately limits our  capability to constrain
 the high-energy spectrum.  
The chance of finding
a source in the PDS beam area can be
evaluated using the HEAO--1 A4 results and 
assuming a $logN-logS$ slope of $-$1.5. 
The HEAO--1 A4 all sky catalog (Levine et al. 1984) lists just
7 high Galactic latitude sources in the 13--80~keV band down to a flux
of 2$\times$10$^{-10}$ \cgs (10mCrab).  Extrapolating this number of 
sources to the  average flux $1.3\times10^{-11}$ \cgs detected
in the energy band 15--100 keV, a chance coincidence rate of $\simeq$ 2 \% 
per target position 
is expected in the PDS at high Galactic latitude
(Elvis et al. 2000).
We also searched  the NED and Simbad catalogs for possible
sources that might contribute to the PDS detection.  The Seyfert 1
galaxy MS 1112.5$+$4059 lies at about 40 arcmin from PG~1115$+$407. Its
X-ray 0.3--3.5 keV flux was about $1.5 \times 10^{-12}$ erg cm$^{-2}$
s$^{-1}$ during the Einstein observation (Stocke et al. 1991).  Moreover, a PSPC
observation of this source shows a steep spectrum ($\alpha$=1.61), with
flux variability of less than 40 \% (Ciliegi \& Maccacaro 1996, 1997).
Unless a large variation (factor of 10) or a sharp hardening
of the spectrum above 3
keV (not usually observed in Seyfert 1 galaxies) occurred, this object cannot be
the responsible for the PDS flux detected.
At about 80 arcmin from the quasar there is also the
Abell cluster A1190, which has a diameter of about 40 arcmin 
putting part of its emission within the PDS field of view (FOV). 
Its 2--10 keV flux is
$3.6\times10^{-12}$ erg cm$^{-2}$ s$^{-1}$ (Ebeling et al. 1996).
Again, unless a very hard component dominates the spectrum above 10
keV, this object cannot be responsible for the PDS flux.  
Three quasars have been recently discovered by Fiore et al. (1999a)
within 6 arcmin of PG~1115$+$407 at redshifts 0.4--1.3. 
Their 2--10 keV flux is however at least 10 times smaller than that
of the PG quasar and therefore they are unlikely to contribute
to the PDS flux, unless their spectrum is strongly inverted. 
\\
In the case of PG~0947$+$396 the quasar KUV 09468$+$3916 lies  about 25
arcmin away. The 0.1--2.4 keV flux of this source is about 30 \% of 
PG~0947$+$396 (Yuan et al. 1998) and again it can not easily be 
responsible of the PDS flux detected.  
\\
We conclude that the PDS flux in both observations
is most likely dominated by the PG quasars and investigate in \S 5 if
the presence of a Compton reflection component could justify the
excess.

\section{0.1--10 keV spectral analysis}

We present in this section the result of the analysis of the
LECS and MECS spectra of the sample quasars, and defer to  \S 5
the inclusion of the PDS data for PG~1115$+$407 and PG~0947$+$396.

 Two sets of fits have been
performed: (1) leaving $N_{\rm H}$ free to vary; and (2) fixing $N_{\rm H}$ to its
Galactic value. 
The best-fit parameters and the $\chi^2$ for each quasar are reported
in Table~\ref{tab3} together with the  fitted fluxes in 
the 0.1--2  and 2--10 keV energy bands,  with the 2--10 keV luminosity, 
and with
the fluxes or 2$\sigma$ upper limits measured by the PDS
in the energy range 15--100 keV
computed extrapolating the LECS-MECS  spectral shapes.  
The $\chi^2$ are generally rather small, not
surprisingly, given the relatively modest statistics. Nevertheless,
analysis of the residuals shows systematic local deviations below 1 keV, 
above 5~keV 
and in the Iron K-$\alpha$ line region, which we discuss in turn.

Positive residuals below 1 keV are evident in  Fig.~\ref{fig1} for six of the
ten sources: PG~0947$+$396, PG~1048$+$342, PG~1115$+$407, PG~1226$+$023, 
PG~1402$+$261  and PG~1626$+$554.  In all these cases, the fits with 
$N_{\rm H}$ free to vary
gives unphysical values, smaller than the Galactic column along the
line of sight. The decrease of the $\chi^2$ between model (1) and model
(2) for the six quasars  is significant
at the 99.8, 93.4, 99.8, $>$99.9, 90.0, and 96.7~\% level 
respectively.
Two of these quasars (PG~1048$+$342 and PG~1626$+$554) also have
a residual background excess (see previous section). However the
excess of LECS counts below 1 keV with respect to the best fit
power-law model is 2--5 times higher than the residual background excess:
$1.5\times10^{-3}$ counts s$^{-1}$ against $3\times10^{-4}$ counts s$^{-1}$ in PG
1626$+$554, $1.0\times10^{-3}$ counts s$^{-1}$ against $5\times10^{-4}$
counts s$^{-1}$ in PG~1048$+$342. We are therefore confident that in all six
quasars the excess of counts below 1 keV is real. 
\\
The case of  PG~1626$+$554 looks rather interesting showing an unusual 
large excess below 0.2 keV; we further discuss this possibility in the next section.

\begin{table*}
\caption{Results of the  fit with a broken power law}
\label{tab4}
\begin{center}
\begin{tabular}{l@{\hspace{0.3cm}}|c@{\hspace{0.5cm}}|c@{\hspace{0.3cm}}cclc}
\hline
\multicolumn{1}{c}  { }      &
\multicolumn{1}{|c} {PSPC}  &
\multicolumn{5}{|c} {BeppoSAX} \\
\multicolumn{1}{c}  {Quasar} &
\multicolumn{1}{|c} {Energy Index} &
\multicolumn{2}{|c} {Energy Index} &
\multicolumn{1}{c} {E$_{br}$} &
\multicolumn{1}{c} {$\chi^2_{\rm bpw}$(dof)} &
\multicolumn{1}{c} {F($\chi^2_{\rm pw},\chi^2_{\rm bpw}$)}\\
                           &  
                           &
\multicolumn{1}{c} {Soft } &
\multicolumn{1}{c} {Hard } &       
\multicolumn{1}{c} {(keV)} &
                           &                          
    \\
\hline
PG~0947$+$396 & 1.51$\pm$0.02 &   1.42$\pm$0.16 & 0.80$\pm$0.08 &
1.09$\pm$0.34 & 48.8 (31) & 0.997\\
PG~1048$+$342 & 1.39$\pm$0.05 &   1.09$\pm$0.16 & 0.79$\pm$0.07 &
1.10$\pm$0.69 & 22.0 (37) & 0.935\\
PG~1115$+$407 & 1.89$\pm$0.04 &   2.15$\pm$0.43 & 1.26$\pm$0.10 &
0.64$\pm$0.32 & 36.9 (32) & 0.998\\
PG~1202$+$281 & 1.22$\pm$0.02 &   0.70$\pm$0.19 & 0.99$\pm$0.06 & 1.0    
      & 26.6 (28) & 0.821\\
PG~1226$+$023$^a$ & 0.94$\pm$0.01 &   1.03$\pm$0.3 & 0.61$\pm$0.01 &
0.55$\pm$0.30 & 253.1 (204) & $>$0.999\\
PG~1352$+$183 & 1.52$\pm$0.03 &   1.29$\pm$0.15 & 1.20$\pm$0.13 & 1.0    
      & 22.3 (20) & 0.390\\
PG~1402$+$261 & 1.93$\pm$0.03 &   1.59$\pm$0.10 & 0.52$\pm$0.30 & 3.4$\pm$0.5  
      & 43.2 (52) & $>$0.999\\
PG~1415$+$451 & 1.74$\pm$0.04 &   1.37$\pm$0.15 & 1.49$\pm$0.16 & 1.0    
      & 16.5 (21) & 0.467\\
PG~1512$+$370 & 1.21$\pm$0.04 &   1.12$\pm$0.21 & 1.12$\pm$0.13 & 1.0    
      & 16.7 (20) & --\\
PG~1626$+$554 & 1.94$\pm$0.04 &   1.52$\pm$0.21 & 1.16$\pm$0.06 &
0.78$\pm$0.46 & 47.0 (41) & 0.903\\
\hline
\end{tabular}
\end{center}

$^a$ The fitting model  includes  an absorbing edge (see text)
\end{table*}


PG~1402$+$26 clearly shows the presence of a hard tail (E $>$5 keV) which is
statistically more significant  than the soft excess (see Fig.~\ref{fig1})
making the $\chi^2$ variation between model (1) and 
model (2) marginally significant (90~\%).
PG~1626$+$554 also shows an excess with respect to a simple power law 
above 5 keV, but, with a lower statistical significance.

We have then fitted the LECS and MECS spectra of the quasars with a
broken power-law model:
\begin{center}
\begin{math}
\begin{array}{llcl}
f_{E} = & e^{-N_{\rm H} \sigma(E)}f_{o} \, E^{-\alpha_{1}} &  & E \leq E_{br} \\
f_{E} = & e^{-N_{\rm H} \sigma(E)}f_{o} \, E_{br}^{-(\alpha_{1}-\alpha_{2})} \, 
E^{-\alpha_{2}} &  & E \geq E_{br} \\
\end{array}
\end{math}
\end{center}
The absorbing column $N_{\rm H}$ has been fixed to the Galactic value. In fact,
leaving it as a free parameter gives values fully compatible with the Galactic
ones but strongly increases  the error on the soft energy index.
The evaluation of the harder spectral index based essentially on MECS
data is not affected by the different source redshifts: the MECS effective area
varies only slowly in the range 2.5--7~keV. 

The
reduction in $\chi^2$ between the broken power-law and the single
power-law fit is significant in PG~0947$+$396 (99.7 \%), PG~1115$+$407
(99.8 \%) and PG~1402$+$261 (99.9~\%), 
again confirming significant curvature in the spectra of
these quasars.  It is marginally significant in PG~1626$+$554 (90.3 \%)
and PG~1048$+$342 (93.5 \%).  In the case of PG~1226$+$023 we have also
introduced in the model an absorbing edge to account for the 0.5--1
keV feature (Grandi et al. 1997).  The improvement in $\chi^2$ between the broken
power-law$+$edge model and the power-law$+$edge model is significant
at the 99.998 \% level.  The parameters of the absorbing edge are
E=0.64$\pm$0.07 keV and $\tau=0.74\pm$0.21, consistent with the
results of Grandi et al. (1997) and Orr et al. (1998).

For the other four quasars the fit gives a $\chi^2$ similar to the
single power-law fit, and the break energy is ill-defined.  To obtain
low and high energy indices to be compared with those of the quasars
with significant curvature we repeated the fits of the PG~1202$+$281, 
PG~1352$+$183, PG~1415$+$451, PG~1512$+$370 spectra fixing the break energy at
1 keV.  We also tried other different values of break energy between
0.5 and 1.5 keV and, as expected, no significant variations in the
fitting parameters and in the $\chi^2$ can be detected.  The
parameters relative to the  fits and the $F$-tests for the
$\chi^2$ variations from the single power-law fits are reported in Table~\ref{tab4}
along with the best fit power-law PSPC energy index.
Note that the break energy in the PG~1402$+$261 fit is significantly 
higher than  found for the others,  as expected from the presence 
of the hard tail. 

Figure~\ref{fig5}  shows the soft energy indices ($\alpha_S$(BSAX))
vs the hard ones ($\alpha_H$); points marked with squares represent 
the six quasars where  spectral curvature may be present.

We plot in Fig.~\ref{fig2}  BeppoSAX soft energy index ($\alpha_S$(BSAX)) as a
function of the PSPC one ($\alpha_S$(PSPC)).  A systematic shift
between the BeppoSAX measurements and the PSPC ones is  evident;
this can be modeled as a constant difference of $\Delta\alpha_E=-0.27\pm0.03$ between the two spectral
indices. 
This difference could be due to the differences between the LECS and
PSPC response functions coupled with low energy spectral steepening of
the intrinsic quasar emission, or it may be due to calibration errors in
one of the two detectors. A more detailed comparison of the PSPC and LECS
results is described in the Appendix.
Figure~\ref{fig6}  shows $\alpha_S$(PSPC) vs $\alpha_H$. 


\setcounter{figure}{1}
\begin{figure}
\centerline{ 
\psfig{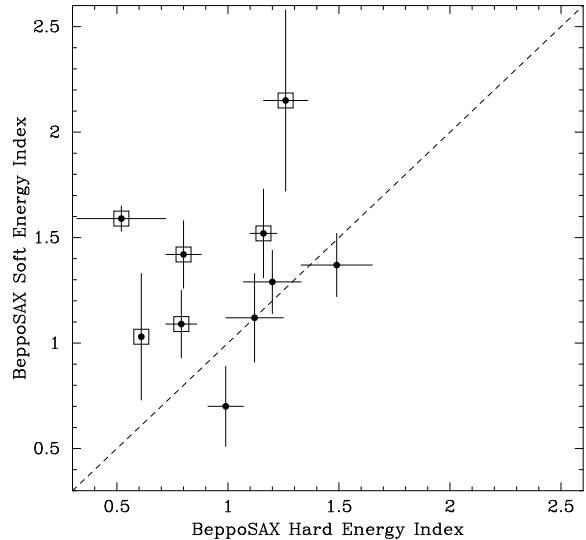}
}
\caption{ 
BeppoSAX soft energy indices $\alpha_S$(BSAX) vs the hard
ones $\alpha_H$ for the observed quasars. Points marked with 
squares represent the six quasars where a
spectral curvature is present  and  
the dashed line identifies the locus of $\alpha_S$(BSAX)=$\alpha_H$} 
\label{fig5}
\end{figure}

\begin{figure}
\centerline{ 
\psfig{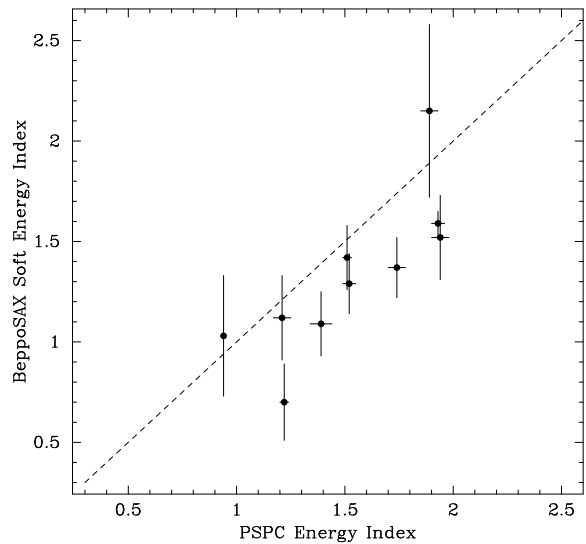}
}
\caption{ 
BeppoSAX soft energy indices $\alpha_S$(BSAX) vs the PSPC ones 
$\alpha_S$(PSPC) for the observed quasars.   
The dashed line identifies the locus of $\alpha_S$(BSAX)=$\alpha_S$(PSPC)}
\label{fig2}
\end{figure}

\begin{figure}
\centerline{ 
\psfig{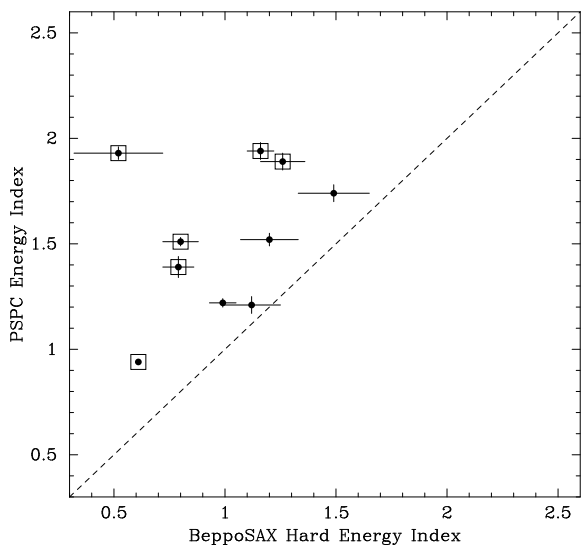}
}
\caption{ 
PSPC energy indices $\alpha_S$(PSPC) vs  BeppoSAX $\alpha_H$ for the 
observed quasars. Points marked with squares represent the six quasars where a
spectral curvature is present  and  
the dashed line identifies the locus of $\alpha_S$(PSPC)=$\alpha_H$}
\label{fig6}
\end{figure}

We also fitted the quasar spectra in which a soft excess is detected
with other composite models:
a double power law of the form $f_{E}=e^{-N_{\rm H} \sigma(E)}(f_{S} \, E^{-\alpha_S}
$+$f_{H} \, E^{-\alpha_H})$, a power law plus blackbody and
a power law plus bremsstrahlung. In all cases the quality of the fits is  
acceptable and does not allow  us to discriminate between models.
Moreover the uncertainties in the soft component parameters are quite
large. Fits with a power law plus bremsstrahlung always give 
large upper limits on the temperatures.
In  Table~\ref{tab5}  
results from the other two models
are reported.
 In particular, we show
the two energy indices and the energy where the two components
have the same flux (E$_c$) for  the double power-law model and the $kT$ and the 
energy index for power law plus blackbody.

\begin{table*}
\caption{Results of composite fits on quasars with a detection of soft excess}
\label{tab5}
\begin{center}
\begin{tabular}{l@{\hspace{0.28cm}}|c@{\hspace{0.28cm}}ccl@{\hspace{0.28cm}}|ccl }
\hline
\multicolumn{1}{c} {Quasar} &
\multicolumn{4}{|c} {Double power law} &
\multicolumn{3}{|c} {Power--law plus Blackbody}\\
\multicolumn{1}{c} { } &
\multicolumn{1}{|c} {$\alpha_S$} &
\multicolumn{1}{c} {$\alpha_H$} &
\multicolumn{1}{c} {E$_c$} &
\multicolumn{1}{c} {$\chi^2$ (dof)} &
\multicolumn{1}{|c} {$kT$} &
\multicolumn{1}{c} {$\alpha_H$} &
\multicolumn{1}{c} {$\chi^2$ (dof)} 
\\
                           &  
\multicolumn{1}{|c} { } &
\multicolumn{1}{c} {} &       
\multicolumn{1}{c} {(keV)} & 
\multicolumn{1}{c} {} &
\multicolumn{1}{|c} {(keV)} &
\multicolumn{2}{c} {}                        
    \\
\hline
 PG~0947$+$396 &    2.41$\pm$1.17 & 0.77$\pm$0.16 & 0.34 & 51.0 (31) &    0.08$\pm$0.02 & 0.86$\pm$0.08 &  51.1 (31)\\
 PG~1048$+$342 &    1.75$\pm$1.95 & 0.73$\pm$0.34 & 0.25 & 22.4 (37) &    0.09$\pm$0.02 & 0.79$\pm$0.07 &  22.1 (37)\\
 PG~1115$+$407 &    3.92$\pm$1.80 & 1.25$\pm$0.11 & 0.27 & 38.8 (32) &    0.05$\pm$0.02 & 1.27$\pm$0.07 &  37.3 (32)\\
 PG~1226$+$023$^a$ & 1.69$\pm$0.61 & 0.58$\pm$0.04 & 0.14 & 252.0 (204) & 0.04$\pm$0.02 & 0.61$\pm$0.01 &  253.4 (204)\\
 PG~1402$+$261 &    1.78$\pm$0.07 & 0.52 & 3.1 & 51.0 (53)              &  $<$0.14      & 1.35$\pm$0.05 &  58.3 (54) \\
 PG~1626$+$554 &    3.73$\pm$3.54 & 1.19$\pm$0.06 & 0.19 & 43.5 (39) &    0.07$\pm$0.02 & 1.19$\pm$0.06 &  43.3 (39)\\
\hline
\end{tabular}
\end{center}
$^a$ The fitting models include an absorbing edge \\
\end{table*}


\subsection{Spectral curvature}

The distribution of the measured spectral indices 
is the result of the convolution of the parent
distribution with the  error distribution.
Assuming that both distributions can be described by a gaussian, it is
possible to deconvolve them  to obtain the best estimate
of the intrinsic dispersion using a
maximum  likelihood technique as explained in Maccacaro et al. (1988).
The intrinsic  dispersions of $\alpha_S$(BSAX), $\alpha_S$(PSPC)
and $\alpha_H$ are then 0.22$\pm$0.14, 0.33$\pm$0.12 and 0.25$\pm$0.17 
respectively, where errors are
at the 90 \% confidence level.
We  find  that  intrinsic
dispersion is present both in the soft and high energy spectral indices.

We have divided the quasar sample into two groups depending on the
$\alpha_S$(BSAX) value: $\alpha_S <$ 1.2 (4 quasars) and $\alpha_S >$1.2
(6 quasars) and computed  the averages 
applying the same maximum  likelihood technique:

\noindent
$\alpha_S$(BSAX)=0.98$\pm$0.13, $\alpha_H$=0.85$\pm$0.18,
$\alpha_S$(PSPC)=1.19$\pm$0.14

\noindent
for the first group and,

\noindent
$\alpha_S$(BSAX)=1.51$\pm$0.08, $\alpha_H$=1.08$\pm$0.20,
$\alpha_S$(PSPC)=1.75$\pm$0.12

\noindent
for the second one. 
Quasars with a steeper soft energy spectrum also tend to
be steeper at high energy, although the two mean high energy indices
differ only at the 68~\% confidence level. High energy observations
for a larger quasar sample
are clearly needed to strengthen this conclusion.

The  residuals from fitting  a simple power law plus Galactic absorption 
to the PG~1626$+$554 spectrum presented in Fig.~\ref{fig1}
seem to show what appears to be an unusually large excess at 0.2 keV.
We investigated this  feature in the PSPC spectrum.
This source has the highest soft X-ray to optical flux ratio in Laor et al. sample (mid panel
of Fig. 3 of Laor et al. 1997). Moreover, the fit with a free $N_{\rm H}$
(Table 3 of that paper) gives an absorption  column lower than the
Galactic one reducing  the $\chi^2$ significantly.
Fitting independently LECS and PSPC spectra in the energy range 0.1--1.5 keV
gives good $\chi^2$ values 
(11.5 for 15 dof in the LECS
and 24.8 for 26 dof in the PSPC) but
incompatible spectral index  1.43$\pm$0.10 and 1.94$\pm$0.04 respectively.
Note however that 
the BeppoSAX spectral index is compatible with
that measured by the PSPC leaving $N_{\rm H}$ as  free parameter (Table 3 
of Laor et al.)
A highly curved soft X-ray spectrum 
could explain  the large discrepancy between the PSPC and the LECS.
An "ultra soft" excess below ~0.3 keV could  be hinted in this object and 
we are probably observing the tail
 of a very steep unusually strong thermal component. 
 Moreover, this "ultra soft" excess could be related to the  unusually flat
optical-to-soft X--ray slope $\alpha_{os}$=-1.139 (Laor et al. 1997).

\section{High energy features}

\subsection{Iron K--lines} 
High energy residuals suggesting the presence of spectral
features  around 5--6 keV (in the observer's frame) have been observed 
in several sources: PG~0947$+$396, PG~1115$+$407 and PG~1352$+$183 
(see Fig.~\ref{fig1}). 

An Iron K--line has previously been detected in PG~1226$+$023 at an
energy compatible with the 
values of many Seyfert 1 galaxies but with a much lower equivalent width (30$\pm$20 eV).

\begin{figure}
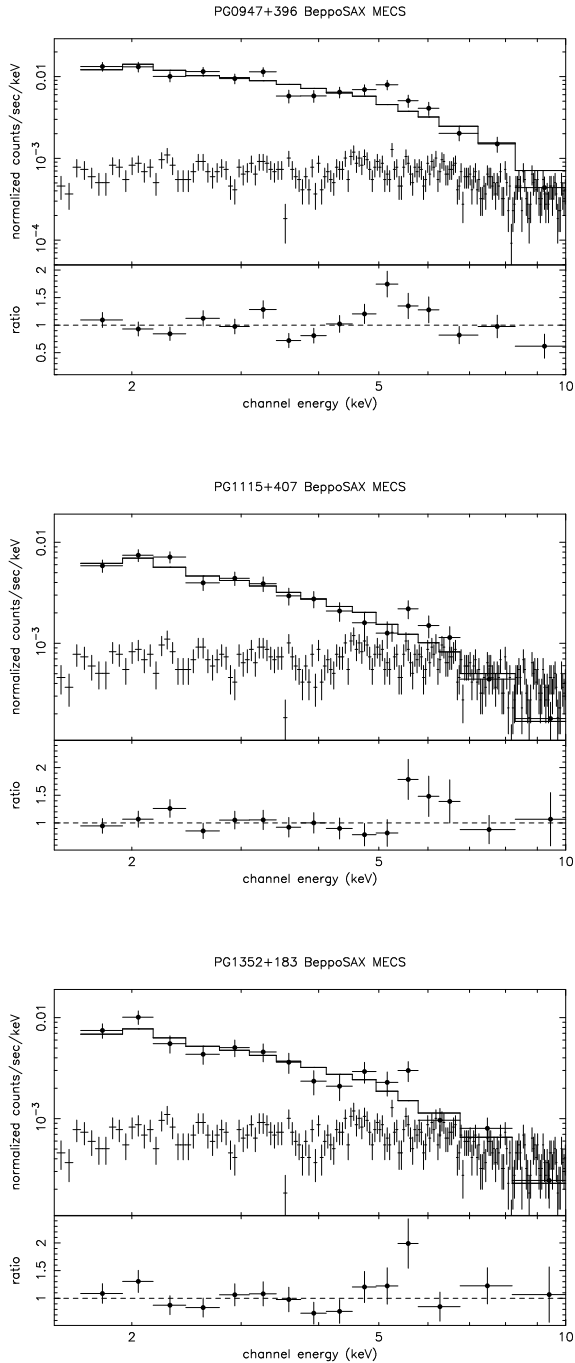

\centerline{
\vbox{ 
\psfig{figure=ms9643_a.f5,width=8.5cm,angle=-90,clip=}
\psfig{figure=ms9643_b.f5,width=8.5cm,angle=-90,clip=}
\psfig{figure=ms9643_c.f5,width=8.5cm,angle=-90,clip=}
}
}
\caption{PG~0947$+$396, PG~1115$+$407 and PG~1352$+$183 MECS spectra 
fitted with a single power law (superimposed as a continuous
curve). For comparison the background
is also shown.  The lower panels represent data--to--model ratios.}
\label{fig8}
\end{figure}

To investigate the statistical significance of the high energy
features, we fitted the MECS spectra of the three quasars with a 
power law and then with a power law plus a narrow line. Results are given in
Table~\ref{tab7}  together with the source redshift.  Figure~\ref{fig8}  shows in the
upper panels the source spectra fitted with a single power law along
with the background and in the lower panel the residuals as
data-to-model ratio.  Note the excess of counts at about 5.8 keV in PG
1115$+$407, 5.2 keV in PG~0947$+$396 and 5.6 keV in PG~1352$+$183. These excesses
 do not
coincide with the largest features in the background (at about 5.0 and
6.0 keV). Moreover the presence of these excesses cannot be justified
by a simple statistical
coincidence. In fact
the probability to get only one of them with the observed significance
by chance is less than 3~\% and looking at a 10 BL Lac sample observed by 
BeppoSAX (Wolter et al. 1998) with similar statistical significance 
we note that no features in the range 5--6 keV are present.

The inclusion of a narrow line at 6.4 keV (quasar frame) is
significant at the 97.7 \% and 96.9 \% (using the $F$-test) in PG~0947$+$396, 
and PG~1352$+$183 respectively. This suggests fluorescence from
cold iron for the origin of the line, similar to what is found in many
local Seyfert 1 galaxies (e.g. Nandra et al. 1997).  The equivalent
widths are however large (670$\pm$170 eV and 760$\pm$460 eV),  compared to those
found in lower redshift Seyfert 1 galaxies
(100--200 eV). In PG~1352$+$183 the uncertainty is large and the EW could be
consistent with that of local Seyfert 1 galaxies.
In PG~0947$+$396 the
statistics  is better, the count rate between 4--6 keV is about ten
times that in the background, and the measured EW is
$>200~eV$ also at the 2.7 $\sigma$ level. 
We therefore investigated whether the high line equivalent  width
is the
results of fitting a simple power-law model to a complex spectrum.
We tried fits with a partial covering model, a broken power law with
a break around 5 keV and a power law plus Compton reflection model
(see also next section), but
in all cases there is no significant reduction in the
$\chi^2$. In all cases the line is still required and its 
equivalent width is never smaller than 400 eV.
Note that this object shows only weak UV absorption (Brandt, Laor \& Wills 2000)
and no strong X-ray absorption is expected to be present 
unless it is  highly variable: the presence of partial covering absorption would
have been difficult to justify.

In PG~1115$+$407 the inclusion of a line gives an energy 6.7$\pm$0.1
keV (probability of 97.0 \% using the $F$-test).  It is however
interesting to note that its energy and equivalent width (580$\pm$280
eV) are similar to those found in Narrow line Seyfert 1 galaxies (TON
S 180; Comastri et al. 1998, Turner et al. 1998) and consistent 
with the EW expected from ionized gas.  Moreover the low and high energy 
slopes are similar to
those of TONS 180, and the $H_{\beta}$ FWHM is only 1720 km
s$^{-1}$, suggesting that PG~1115$+$407 may be  a narrow--line Seyfert 1 
galaxy at higher redshift.

Upper limits at 90 \%  confidence level 
on  narrow  line equivalent widths  at 6.4 and 6.7  keV 
in all other quasars 
are also computed:  values range between 170 and 770 eV.

\subsection{High Energy Spectra}
PG~0947$+$396  and PG~1115$+$407  show a  signal at $>2\sigma$ in the PDS
that can likely be attributed to the two quasars.
We therefore investigate whether
a Compton reflection model from a neutral slab (model {\bf pexrav} in XSPEC)
 could justify the two detections.
 In both cases the PDS detection was always above the model
despite the best fit normalization for the
Compton reflection component was at the  limit of the 
allowed range (R$_{cp}$=5 is the limit 
for the reflection component scaling factor).
The reduction in $\chi^2$ was never significant. 

We also tried to fit the hard excess in PG~1402$+$26 and PG~1626$+$554  
with the Compton reflection component  including 
 the PDS upper limits. The disk inclination 
has been arbitrarily
fixed to 45\degr and a narrow line at fixed energy has been included into the model.
For  PG~1402$+$26 we get
a spectral index of 1.62$\pm$0.07, a  scaling factor for the reflection
component (R$_{cp}$) of 3.8$\pm$2.2 and a 2 $\sigma$ upper limit of 350 eV for the 
EW of the 6.4 keV iron line.
The $\chi^2$ for this model (42.8 for 49 dof) decreases with
respect to the value obtained for the simple  power law in the 0.1--100~keV range
(45.9 for 50 dof) with a significance of  96~\%.
\\
For PG~1626$+$554 we get a spectral index of 1.38$\pm$0.06, a  scaling factor of
2.2$\pm$1.1 and a 2 $\sigma$ upper limit of 400 eV for the EW of the 
6.4 keV iron line. The reduction in $\chi^2$ with respect to the simple
 power law 
(49.3 for 46 dof with respect to  53.1 for 47 dof)
corresponds to  a  significance level of 93 \%.

\begin{table*}
\caption{Results of the MECS data fit with a single power law plus line}
\label{tab7}
\begin{center}
\begin{tabular}{lcccccl}
\hline
\multicolumn{1}{c} {Quasar} &
\multicolumn{1}{c} {Fit$^a$} &
\multicolumn{1}{c} {z} &
\multicolumn{1}{c} {Energy Index} &
\multicolumn{1}{c} {E$_{line}$} &
\multicolumn{1}{c} {EW} &
\multicolumn{1}{c} {$\chi^2$ (dof)} \\
       &   &  &   & \multicolumn{1}{c} {(keV)} & \multicolumn{1}{c} {(eV)} &    \\
\hline
PG~0947$+$396 & 1 & 0.206 & 0.85$\pm$0.09 &     --        & --           
& 32.5 (14) \\
            & 2 &       & 0.95$\pm$0.10 & 6.35$\pm$0.13 & 670$\pm$170  
& 17.3 (12) \\
            
                      
PG~1115$+$407 & 1 & 0.154 & 1.29$\pm$0.12 & --            & --           
& 14.1 (14) \\
            & 2 &       & 1.40$\pm$0.13 & 6.69$\pm$0.11 & 580$\pm$280  
& 7.8  (12) \\
            

PG~1226$+$023 & 1 & 0.158 & 0.60$\pm$0.06 &  --           &   --         
& 233.9 (182) \\
            & 2 &       & 0.60$\pm$0.06 & 6.21$\pm$0.09 & 30$\pm$20    
& 219.1 (180) \\
            
PG~1352$+$183 & 1 & 0.158 & 1.16$\pm$0.14 &  --           &   --         
& 14.1 (13) \\
              & 2 &       & 1.30$\pm$0.16 & 6.43$\pm$0.16 & 760$\pm$460  
& 8.4 (11)  \\
\hline


            
               

\end{tabular}
\end{center}

$^a$ Fit 1:  power law plus  Galactic absorption; 
Fit 2:  same as Fit 1 plus a narrow line with energy relative to the quasar frame. 
\end{table*}


\section{Conclusions}

We presented the spectral analysis of the BeppoSAX observations of
10 PG  quasars selected from the Laor et al. (1997)
sample. The main results can be  summarized as follows:
\begin{itemize}

\item[$\bullet$]
Together with PG~1226$+$023
a positive detection  of the continuum in the 15--100 keV energy range has been found 
in two more quasars: PG~0947$+$396 and PG~1115$+$407. 
However, a possible contamination from hard serendipitous sources 
in the PDS field of view cannot be ruled out with the present data.

\item[$\bullet$]
The distribution of the 2--10 keV power law energy indices 
is similar to that observed in other quasars samples. The
dispersion around the average value of $\alpha_E = 1.0\pm 0.3$  reflects
the large spread of the best--fit values over the range 
0.5 $< \alpha_E <$ 1.5.

\item[$\bullet$]
 No intrinsic absorption has been detected in any of the objects
since the absorbing columns are always compatible with the Galactic values.

\item[$\bullet$]
Significant spectral curvature is present in the BeppoSAX spectra
of most of the quasars and is related to the statistics
of the observed spectrum: sources with the same energy index
over the 0.1--10 keV band are fainter.
The average curvature can be parameterized  
as a spectral flattening by $\Delta\alpha\sim0.5\pm0.2$ 
towards high energies requiring two component models.
The exact spectral shape and intensity of these components
vary from object to object.
In a few cases the curvature is due to a strong 
``soft excess" below about 1 keV, while for 
PG 1626$+$554,  independent LECS--PSPC fits suggest the presence 
of an "ultra soft excess" (below ~0.3 keV).
The curvature in PG~1402$+$26  
is mainly due to a hard tail above $\sim$ 5 keV 
rather than a ``soft excess"; also PG~1626$+$554  shows
evidence for a hard tail.
A spectral hardening at even higher energies ($>$ 10 keV) 
could be present in the two quasars detected in the PDS band.

\item[$\bullet$]
The origin of the spectral curvature is  likely  due to the
combined effect of thermal emission from an accretion disk 
peaking in the far UV and the onset of a Compton reflection
component at high energies.
The addition of these components  does provide a 
better description of the observed spectra.
We also note that quasars with a steeper 0.1--2 keV spectrum 
tend to be steeper also in the 2--10 keV band, although the effect is detected 
only at the 68 \% confidence level.

\item[$\bullet$]
Iron K$_{\alpha}$ lines are detected in 4  quasars.
The rest frame line energy (6.7 keV) and equivalent width (580 eV) 
of PG~1115$+$407 are  consistent with those found in
a few low redshift narrow--line Seyfert 1 galaxies (Comastri et al. 1998,
Turner et al. 1998, Vaughan et al. 1999, Leighly 1999). The contemporaneous presence
of a steep 0.1--10 keV continuum and narrow H$\beta$ line 
allows  us to classify this object as a relatively high redshift NLSy1.

For the optically broad lined quasar PG~0947$+$396, 
the rest frame line energy of 6.4 keV is 
similar to that found in many local Seyfert 1 galaxies, suggesting 
fluorescence from cold iron. The line equivalent width 
(670 eV) is however higher (at the $2-3 \sigma$ level) than that 
usually found in Seyfert 1 galaxies. We investigate the possibility that 
the high EW of  the line, as well as an excess of counts detected
in the PDS, are 
the result of fitting a complex continuum with a simple power-law model.
We tried several other models 
(the inclusion
of a Compton reflection component or of a thick and partial covering
absorber) but in all cases the line is still required and 
its EW is never smaller than 400 eV.

A line has also been detected in PG~1352$+$183 and PG~1226$+$023.
In the first case the line energy and intensity are compatible with the 
values detected in many Seyfert galaxies. 
In the second case it could be witnessing a Seyfert-like spectrum
diluted in the jet emission (Grandi et al. 1997,  Haardt et al. 1998).

The detection of significant Fe K--shell emission
in 3 radio--quiet quasars with 2--10 keV luminosities 
in the range 1--5 $\times$ 10$^{44}$ erg s$^{-1}$ seems to be inconsistent
with the trend seen in other radio--quiet AGN (Iwasawa \& Taniguchi 1993;
Nandra et al. 1997, George et al. 2000), where iron lines are 
more frequently found in lower luminosity objects. 

\item[$\bullet$]
A detailed comparison  with the recent ASCA observations
of a larger sample of radio--quiet quasars (George et al. 2000) is not 
possible owing to the different responses, sensitivities and energy 
ranges covered by ASCA and BeppoSAX.
We note however that there is  a good agreement between the two samples
especially for what  concerns the average 2--10 keV slope and
intrinsic dispersion and the presence of curved convex spectra in most of the 
objects. The detection with BeppoSAX of a strong iron line in a 
few relatively high luminosity objects 
deserves further investigation. The foreseen XMM-Newton quasar surveys  
will most likely settle several open issues.

\end{itemize}

\begin{acknowledgements}
TM and FF thank A. Wolter for the useful discussion on the LECS--PSPC
intercalibration and P.Ciliegi for the several comments and suggestions 
to the paper. 
WNB acknowledges the support of NASA LTSA grant NAG5-8107, and
  AC acknowledges the support of Italian Space Agency contract 
  ASI-ARS-98-119 and MURST grant Cofin-98-02-32. 
\end{acknowledgements}

\appendix

\section{ Comparison with PSPC}

\begin{figure}
\centerline{ 
\psfig{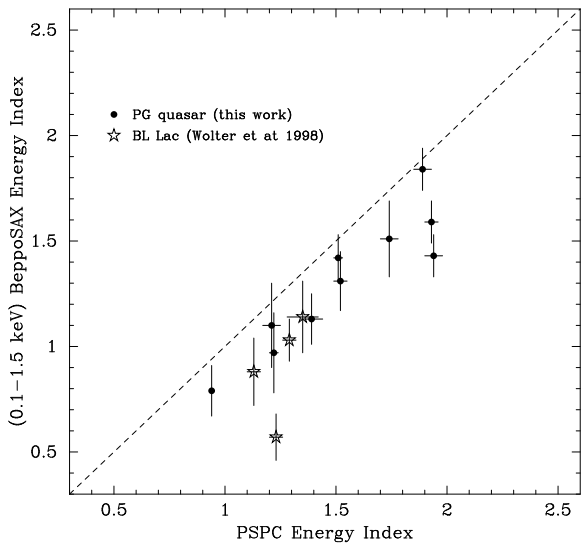}
}
\caption{ 
LECS energy index $\alpha_{LECS}$ detected in the energy range 
0.1--1.5 keV vs the PSPC energy index $\alpha_S$(PSPC) for the observed 
quasars plus four BL Lacs objects
from the sample of Wolter et al. (1998)
}
\label{fig3}
\end{figure}

The discrepancy between the BeppoSAX soft spectral index and the PSPC
one can be due to systematic errors in the instrument intercalibration.
To investigate this discrepancy we fitted the LECS spectra in
the energy range 0.1--1.5 keV with a power law plus Galactic
absorption.  Considering the different response curves of the two
instruments around 1 keV (the LECS effective area increases from 1 to
2 keV, the PSPC area decreases in the same range) this is the LECS
range which is more appropriate for a quantitative comparison with the
PSPC. For the PG~1226$+$023 and PG~1202$+$281 fits we have added an edge at 
0.6 keV and 0.8 keV respectively in the
model.  The resulting energy indices ($\alpha_{LECS}$) are shown in
Fig.~\ref{fig3}  as a function of the PSPC ones, together with the
$\alpha_{LECS}= \alpha_S$(PSPC) line.  To increase the sample we have
added four BL Lacertae objects observed by BeppoSAX  (Wolter et
al. 1998). BL Lacs objects usually do not show large
neutral or ionized low energy absorption (unlike Seyfert 1 and 2 galaxies)
The four BL Lacs were selected from the ten of the Wolter et
al. sample, excluding the objects with a complex (or variable)
soft X-ray spectrum.

A systematic difference between the LECS and PSPC spectral indices is
again evident.  The BeppoSAX soft spectral  indices vs the PSPC ones can
be well fitted by a line with slope {\it m}=0.97$\pm$0.10 and offset 
$\Delta\alpha_E$=-0.23$\pm$0.16. Fixing {\it m} to 1 
gives a
constant difference between the two detectors measurements of
$\Delta\alpha_E=-0.27\pm0.03$. Excluding the most discrepant
point makes a marginal difference.

This discrepancy is smaller than that found between the PSPC and the
Einstein IPC ($\Delta\alpha$=0.5; Fiore et al. 1994) 
and the one between the PSPC 
and the ASCA SIS ($\Delta\alpha=0.4$; Iwasawa et al. 1999). 
In these latter two cases part of the difference
can be due to the fact that the common energy range between these
instruments is not large.  Conversely, the energy range shared by the PSPC and
the LECS is wide (0.1--2 keV), and, in particular, both instruments have
good sensitivity below the Carbon edge at 0.28 keV. Their comparison
is therefore more straightforward. However, if quasar spectra deviate from the
simple power law, a small difference in the spectral index measured by 
the two instruments, that look at different average energies, could be possible.
In order to check this effect
 we have fitted
the LECS and PSPC spectra of  the quasars with the best photon  statistics
simultaneously with curved models (broken power law and the power law 
with energy index variable with energy presented in Laor et al 1994), allowing
for a different normalization in the two instruments and 
letting the model parameters linked together.

The discrepancy in the estimated spectral index of about 0.2
could result from the differences 
in instrument response functions coupled with an energy dependent spectral slope
at 0.1--2 keV (with a reasonable curvature) in about half of the cases. 
It is unclear then that the PSPC 
and BeppoSAX results are always inconsistent.

\begin{figure}
\centerline{ 
\psfig{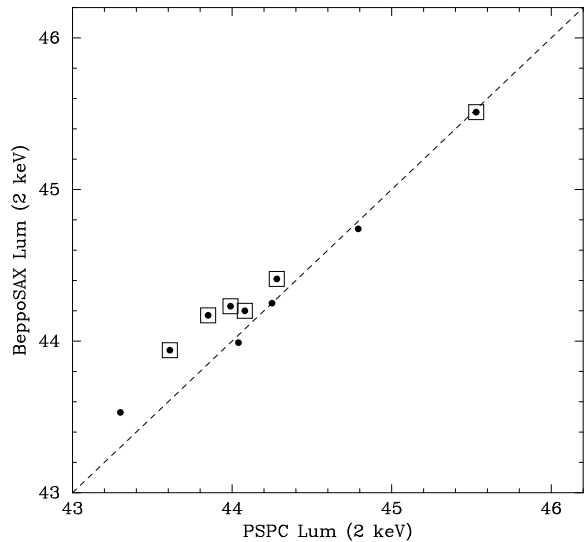}
}
\caption{ 
The 2 keV luminosity estimated from the BeppoSAX observations 
as a function of the PSPC 2 keV luminosity. Values are expressed in
unit of log (erg s$^{-1}$).Points marked with 
squares represent the six quasars where a
spectral curvature is present
}
\label{fig4}
\end{figure}

Figure~\ref{fig4}  shows the BeppoSAX 2 keV luminosity 
as a function of the PSPC 2 keV luminosity.
Six quasars have BeppoSAX luminosities slightly higher than PSPC luminosity:
the BeppoSAX average luminosity is 30 \% higher than the PSPC one.
Variability can certainly play a role in this difference, given the
small sample of objects. However we note that
2 keV are the limit of the PSPC band and therefore a small uncertainty 
on the PSPC spectral shape could translate in a relatively large error on the 
2 keV flux and luminosity. 
The slope of the correlation in Fig.~\ref{fig4}  is consistent with 1, indicating that
this possible shift should not alter the results of the correlations
between luminosity and other quasar properties discussed in Laor et al. 
(1997).

\end{document}